\documentclass[journal]{new-aiaa}
\usepackage[utf8]{inputenc}
\usepackage{textcomp}

\usepackage{graphicx}
\usepackage{amsmath}
\usepackage[version=4]{mhchem}
\usepackage{siunitx}
\usepackage{longtable,tabularx}
\usepackage{bm}
\usepackage{cuted}
\usepackage{subcaption}
\usepackage{lineno}
\usepackage{setspace}
\usepackage{placeins}
\hypersetup{hidelinks}
\title{Approximating Anisotropic Solar Radiation Pressure}

\author{Yasuhiro Yoshimura\footnote{Assistant Professor, Department of Aeronautics and Astronautics, 744 Motooka, Nishi-ku, Fukuoka, Japan; Corresponding Author; y.yoshimura.a64@m.kyushu-u.ac.jp}
and Toshiya Hanada\footnote{Professor, Department of Aeronautics and Astronautics, 744 Motooka, Nishi-ku, Fukuoka, Japan. Senior Member AIAA}}
\affil{Kyushu University, Fukuoka 819-0395, Japan}

\begin{document}

\maketitle

\begin{abstract}
  Solar radiation pressure (SRP) significantly influences satellite dynamics in high-altitude orbits, depending not only on the satellite's trajectory and attitude but also on its surface optical characteristics.
  Metallic surfaces, such as solar cells with periodic metal bars, exhibit anisotropic reflection that produces an anisotropic SRP.
  The difficulty in dealing with anisotropic SRP arises because anisotropic SRP does not have a general analytical solution, and accurate computation therefore requires numerical integration. An analytical formulation of anisotropic SRP is useful for stability analysis and controller design for satellites that exploit SRP.
  To this end, this paper derives an analytical approximation method to calculate the anisotropic SRP modeled with the Ashikhmin--Shirley model.
  The anisotropy is approximated using multiple spherical Gaussians. Because the product of spherical Gaussians is also a spherical Gaussian, anisotropic reflection can be expressed in an analytically tractable form. The approximation accuracy of the anisotropic SRP is numerically verified for two cases: weak anisotropy and strong anisotropy.
\end{abstract}

{\singlespacing\footnotesize\noindent
This is the author's accepted manuscript of the article published in Journal of Guidance, Control, and Dynamics,
\href{https://doi.org/10.2514/1.G010259}{\mbox{https://doi.org/10.2514/1.G010259}}.
Copyright~\copyright~2026 by Yasuhiro Yoshimura and Toshiya Hanada.
Published by the American Institute of Aeronautics and Astronautics, Inc., with~permission.\par}

\section{Introduction} \label{intro}
Solar radiation pressure (SRP) is one of the dominant perturbations acting on satellites in high-altitude orbits, and accurate SRP modeling is essential for both mission design and precise orbit determination.
SRP models fall into three families: empirical, semi-analytical, and analytical.
Empirical models fit simple functional forms to tracking data for a specific spacecraft~\cite{beutler, Montenbruck2017, Duan2021, springer2000modeling, arnold2015}, where the SRP is typically expressed with trigonometric functions in a Sun-oriented frame and the parameters are estimated from observations.
Semi-analytical models estimate reflection and scaling parameters on top of a detailed spacecraft shape. For example, Guo et al.~\cite{guo} consider self-shadowing of the antenna, and Wang et al.~\cite{wangSRP} focus on thermal radiation pressure.
Analytical models, on which this paper focuses, derive the SRP from the surface geometry and optical properties of each facet.
Facet-by-facet evaluation is expensive, and Sochacki et al.~\cite{Sochacki.2024} compare interpolation techniques to reduce the computational cost.
Any analytical improvement that preserves a compact closed form directly benefits solar-sail trajectory design, attitude dynamics, and control~\cite{Quarta2022, caruso, oguri, ono2018, Ji2021, Benson2021a, takao}, as well as the semi-analytical models that build on analytical formulations.

The conventional analytical formulation assumes a Lambertian diffuse component and a perfect mirror-like specular component~\cite{mcinnes2013}.
This idealization gives a closed-form SRP but cannot capture the reflection behavior of real spacecraft surfaces.
Metallic materials~\cite{Ngan.2004} and solar cells~\cite{small} in particular exhibit anisotropic reflection: the specular peak is shifted away from the mirror direction, and its shape depends on the direction of the metal grid lines on the cell.
Because solar cells often cover a large fraction of a spacecraft surface, capturing this anisotropic behavior is important for high-fidelity SRP modeling.

Physics-based bidirectional reflectance distribution functions (BRDFs) from computer graphics, such as the Cook--Torrance model~\cite{cook} and the Ashikhmin--Shirley model~\cite{ashikhmin2000}, describe these phenomena realistically but do not admit closed-form SRP integrals.
As a result, only a few studies have applied them to SRP.
Wetterer et al.~\cite{wetterer2014} fit correction coefficients to the difference between the detailed and the conventional SRP, but their fit assumes isotropic reflection and therefore cannot describe the anisotropic Ashikhmin--Shirley model.
Yoshimura et al.~\cite{yoshimura2022} derive an analytical approximation for the isotropic Cook--Torrance SRP. Their general incidence formulation sums two analytical vector moments over each angular subdomain of a piecewise linear SG approximation.

This paper proposes a compact analytical approximation of the SRP that (1) handles anisotropic reflection directly through the Ashikhmin--Shirley model, (2) reduces to a simple form for the isotropic Cook--Torrance model as a special case, and (3) yields a closed-form expression suitable for stability analysis and control design.
The key idea is to approximate the BRDF as a sum of spherical Gaussians.
Because the product of two spherical Gaussians is itself a spherical Gaussian and the integral of a spherical Gaussian over the unit sphere has a closed form, the resulting SRP integral can be evaluated analytically for arbitrary lobe directions.
Compared with~\cite{yoshimura2022}, the formulation in this paper is expressed as a linear combination of a small number of spherical Gaussians and each of them contributes a single vector term, which makes the formulation simpler to analyze and directly applicable to the anisotropic case.

This paper is organized as follows. Section~\ref{sec:pre} introduces the reflection models considered in this study. The general SRP formulation and spherical Gaussian are also described. Sections~\ref{sec:CT} and~\ref{sec:AS} provide the proposed approximation method for the Cook--Torrance model and the Ashikhmin--Shirley model, respectively.
In Section~\ref{sec:verify}, the proposed method is numerically verified, and Section~\ref{sec:conc} concludes this paper.

\section{Preliminaries}\label{sec:pre}
This section reviews the bidirectional reflectance distribution function (BRDF) and the two reflection models used in this study, the general formulation of the SRP, and the basic properties of spherical Gaussians, which provide the foundation for the proposed approximation method.

\subsection{Bidirectional Reflectance Distribution Function}
\begin{figure}[tb]
  \centering
  \includegraphics{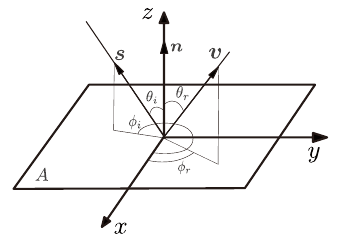}
  \caption{Geometric relation of light reflection on a facet.}
  \label{fig:facet}
\end{figure}

Light reflectance is expressed with a bidirectional reflectance distribution function (BRDF) $f_{r}$~\cite{nicodemus1977} defined as
\begin{align}
  f_{r}(\bm{s}, \bm{v}) = \frac{L_{r}(\bm{v})}{E_{i}(\bm{s})}
\end{align}
where $E_{i}(\bm{s})$ is the irradiance arriving from the incident direction $\bm{s}$, and $L_{r}(\bm{v})$ is the radiance reflected in the direction $\bm{v}$.
The geometric relation of light reflection on a small facet is illustrated in Fig.~\ref{fig:facet}. The facet is defined with the facet-fixed reference frame, where the facet spans the $x-y$ plane and the $z$ axis is along the normal vector $\bm{n}$ of the facet. The unit vectors $\bm{s}$ and $\bm{v}$ represent the directional vector of the Sun and the reference vector, respectively. Their azimuth and polar angles are expressed by $\left(\phi_{i},\theta_{i}\right)$ and $\left(\phi_{r},\theta_{r}\right)$ and $\bm{s}$ and $\bm{v}$ are written as
\begin{align}
  \bm{s} & = \begin{bmatrix}
               \sin{\theta_{i}} \cos{\phi_{i}} \\
               \sin{\theta_{i}} \sin{\phi_{i}} \\
               \cos{\theta_{i}}
             \end{bmatrix}   \\
  \bm{v} & = \begin{bmatrix}
               \sin{\theta_{r}} \cos{\phi_{r}} \\
               \sin{\theta_{r}} \sin{\phi_{r}} \\
               \cos{\theta_{r}}
             \end{bmatrix}
\end{align}
The bisector between $\bm{s}$ and $\bm{v}$ is often used to simplify the reflection model and is defined as follows.
\begin{align}
  \bm{h} = \frac{\bm{s}+\bm{v}}{\|\bm{s}+\bm{v}\|} =\begin{bmatrix}
                                                      \sin{\theta_{h}} \cos{\phi_{h}} \\
                                                      \sin{\theta_{h}} \sin{\phi_{h}} \\
                                                      \cos{\theta_{h}}
                                                    \end{bmatrix} \label{eq:bisector}
\end{align}
where $\phi_{h}$ and $\theta_{h}$ are the azimuth and polar angles of the bisector, respectively.

The BRDF can be decomposed into a diffuse component and a specular component as
\begin{align}
  f_{r} = c_{d} + c_{s} \label{eq:cdcs}
\end{align}
where $c_{d}$ is the diffuse reflection term and $c_{s}$ is the specular reflection term.
Light reflectance has been extensively studied in computer graphics, and many BRDFs~\cite{overviewBRDF} have been proposed to describe realistic scenes.
A simple BRDF combines Lambertian diffusion and perfect mirror-like specularity~\cite{nicodemus1977} as
\begin{align}
  c_{d} & = \frac{\rho_{d}}{\pi} \label{eq:simpleBRDFcd}                                                                            \\
  c_{s} & = 2\rho_{s}\delta(\sin^{2}{\theta_{r}}-\sin^{2}{\theta_{i}})\delta\left(\phi_{i} - \phi_{r} \pm \pi \right) \label{eq:simpleBRDF}
\end{align}
where $\rho_{d}$ is the diffuse reflectance, $\rho_{s}$ is the specular reflectance, and $\delta\left(\cdot\right)$ is the Dirac delta function.
Lambertian diffusion means that incident light is equally reflected in all directions, whereas
specular reflection, as formulated with the Dirac delta function, occurs only in the direction of $\theta_{r}=\theta_{i}$ and $\phi_{r} = \phi_{i} \pm \pi$. This perfect specularity geometrically means that the bisector of $\bm{s}$ and $\bm{v}$ corresponds to the normal vector of the facet $\bm{n}$.

This paper considers two reflection models: the Cook--Torrance model~\cite{cook} and the Ashikhmin--Shirley model~\cite{ashikhmin2000}.
The Cook--Torrance model is an isotropic BRDF whose diffuse term is Lambertian, equivalent to that in Eq.~\eqref{eq:simpleBRDFcd}. The specular term is written as
\begin{align}
  c_{s, {\rm CT}} & = \frac{D_{{\rm CT}}G_{\rm CT}(\bm{v}) F_{{\rm CT}}(\bm{v})}{4(\bm{n}^{T}\bm{s})(\bm{n}^{T}\bm{v})} \label{eq:csCT}
\end{align}
where $D_{\rm CT}$ is the normal distribution function (NDF) describing the statistical distribution of microfacet orientations, $G_{\rm CT}$ is the geometric attenuation function accounting for self-shadowing and masking among microfacets, and $F_{\rm CT}$ is the Fresnel reflection that models the angular dependence of the reflection coefficient.
These are defined as
\begin{align}
  D_{\rm CT}   & = \frac{1}{\pi m^{2}\cos^{4}{\theta_{h}}}e^{-(\tan{\theta_{h}}/{m})^{2}} \label{eq:NDF}                                                                                                                                    \\
  G_{\rm CT}(\bm{v}) & = {\rm min}\left\{ 1, \frac{2(\bm{n}^{T}\bm{h})(\bm{n}^{T}\bm{v})}{\bm{v}^{T}\bm{h}}, \frac{2(\bm{n}^{T}\bm{h})(\bm{n}^{T}\bm{s})}{\bm{v}^{T}\bm{h}}\right\}                                                               \\
  F_{{\rm CT}} & = \frac{(g-\bm{v}^{T}\bm{h})^{2}}{2(g+\bm{v}^{T}\bm{h})^{2}}\left\{1+\frac{\left[\bm{v}^{T}\bm{h}\left(g+\bm{v}^{T}\bm{h}\right)-1\right]^{2}}{\left[\bm{v}^{T}\bm{h}\left(g-\bm{v}^{T}\bm{h}\right)+1\right]^{2}}\right\}
\end{align}
where
\begin{align}
  g^{2}       & = n_{\rm ref}^{2}+(\bm{v}^{T}\bm{h})^{2}-1            \\
  n_{\rm ref} & = \frac{1+\sqrt{F_{0}}}{1-\sqrt{F_{0}}} \label{eq:F0}
\end{align}
The NDF in Eq.~\eqref{eq:NDF} is the Beckmann distribution function and $m$ is the roughness parameter.
Following the convention of the projected-area normalization by Walter et al.~\cite{walter2007}, the NDF is normalized as
$\int_{\Omega^{+}}D_{\rm CT}(\bm{h})(\bm{n}^{T}\bm{h})\,\mathrm{d}\omega_{h}=1$,
where $\mathrm{d}\omega_{h}$ is the solid-angle element for the microfacet normal.
With this convention, the denominator in Eq.~\eqref{eq:csCT} contains the factor 4, whereas the original Cook--Torrance formulation uses a different normalization of the NDF and contains the factor $\pi$.
In Eq.~\eqref{eq:F0}, $F_{0}$ is the Fresnel reflectance at normal incidence.
For convenience in the SRP approximation derived later, an auxiliary function $M(\bm{v})$ is introduced that lumps the geometric attenuation, the Fresnel reflection, and the cosine factors in the denominator as
\begin{align}
  M(\bm{v}) = \frac{G_{\rm CT}(\bm{v})F_{\rm CT}(\bm{v})}{4 (\bm{n}^{T}\bm{s})(\bm{n}^{T}\bm{v})} \label{eq:Mv}
\end{align}
This isolates all the dependence on $\bm{v}$ that does not come from the NDF, and it is used in Section~\ref{sec:CT} to derive the analytical SRP approximation with the Cook--Torrance model.

The Ashikhmin--Shirley model~\cite{ashikhmin2000}, also known as the anisotropic Phong model, uses the following diffuse and specular terms:
\begin{align}
  c_{d,{\rm AS}} & = \frac{28\rho_{d}}{23 \pi} \left(1-F_{0}\right) \left[1-\left(1-\frac{\bm{n}^{T}\bm{s}}{2}\right)^{5}\right]\left[1-\left(1-\frac{\bm{n}^{T}\bm{v}}{2}\right)^{5}\right] \label{eq:AScd} \\
  c_{s,{\rm AS}} & = \frac{\sqrt{\left(n_{u}+1\right)\left(n_{v}+1\right)}}{8\pi} \frac{F_{{\rm AS}}}{(\bm{v}^{T}\bm{h}){\rm max}(\bm{n}^{T}\bm{s},\bm{n}^{T}\bm{v})} D_{{\rm AS}}(\bm{h}) \label{eq:AScs}
\end{align}
where
\begin{align}
  F_{{\rm AS}}         & = F_{0} + (1-F_{0}) (1 - \bm{v}^{T}\bm{h})^{5}                                        \\
  D_{{\rm AS}}(\bm{h}) & = (\bm{n}^{T}\bm{h})^{n_u\cos^{2}{\phi_{h}} + n_{v}\sin^{2}{\phi_{h}}} \label{eq:Das}
\end{align}
The anisotropy is defined with $n_{u}$ and $n_{v}$, which represent the specular lobes along the $x$ and $y$ axes in the facet-fixed frame, respectively.
If $n_{u}=n_{v}$, the reflection becomes isotropic.
It is noted that the diffuse term of the Ashikhmin--Shirley model is not Lambertian, as described in Eq.~\eqref{eq:AScd}.

\subsection{Solar Radiation Pressure}
SRP is the force caused by photon impingement and its reflection on a satellite surface. The SRP generated on a facet is formulated using a BRDF as~\cite{wetterer2014}
\begin{align}
  {\bm{f}_{{\rm{SRP}}}} & = -\int_{0}^{\infty} \frac{(\bm{n}^{T}\bm{s}) F_{s}\left(\lambda_{w}\right) A}{c r^{2}_{\rm AU}}\left[{\bm{s}}+\iint (\bm{n}^{T}\bm{v}) f_{r}(\bm{s},\bm{v}) \sin{\theta_{r}}{\bm{v}}\mathrm{d}\theta_{r}\mathrm{d}\phi_{r}\right] \mathrm{d}\lambda_{w} \\
                        & = -\frac{S_{0}}{cr^{2}_{\rm AU}}(\bm{n}^{T}\bm{s}) A\left({\bm{s}}+\iint (\bm{n}^{T}\bm{v}) f_{r}(\bm{s},\bm{v}) \sin{\theta_{r}}{\bm{v}}\mathrm{d}\theta_{r}\mathrm{d}\phi_{r}\right)\label{eq:fSRP2}
\end{align}
where $F_{s}(\lambda_{w})$ is the solar flux, $A$ is the facet area, $c$ is the speed of light, $\lambda_{w}$ is the wavelength, $S_{0}$ is the solar constant, and $r_{\rm AU}$ is the distance between the satellite and the Sun in astronomical units.
The BRDF is assumed to be independent of the wavelength $\lambda_{w}$, which allows the integration with respect to $\lambda_{w}$ to be separated from the directional integration.
Substituting a specific BRDF into Eq.~\eqref{eq:fSRP2} yields the SRP for a facet characterized by the reflection properties of the BRDF.

As mentioned above, a conventional SRP formulation uses the simple BRDF in Eqs.~\eqref{eq:simpleBRDFcd} and~\eqref{eq:simpleBRDF}, which assumes Lambertian diffusion and perfect specularity.
This is because such a simple BRDF enables analytical integration of Eq.~\eqref{eq:fSRP2}.
Substitution of Eqs.~\eqref{eq:simpleBRDFcd} and~\eqref{eq:simpleBRDF} into Eq.~\eqref{eq:fSRP2} gives
\begin{align}
  {\bm{f}_{\rm{SRP}}} & = -\frac{S_{0}}{cr^{2}_{\rm AU}}(\bm{n}^{T}\bm{s}) A \left\{\bm{s} + \iint \left\{ \frac{\rho_{d}}{\pi}
  + \left[2\rho_{s}\delta(\sin^{2}{\theta_{r}}-\sin^{2}{\theta_{i}})\delta\left(\phi_{i}-\phi_{r}\pm \pi\right)\right] \right\} (\bm{n}^{T}\bm{v})\sin{\theta_{r}}{\bm{v}}\mathrm{d}\theta_{r}\mathrm{d}\phi_{r} \right\} \label{eq:fSRPbrdf}
\end{align}
The integrations of the diffuse and specular terms are analytically obtained as
\begin{align}
   & \int_{0}^{2\pi}\int_{0}^{\frac{\pi}{2}}\frac{\rho_{d}}{\pi}(\bm{n}^{T}\bm{v})\sin{\theta_{r}}\bm{v}\mathrm{d}\theta_{r}\mathrm{d}\phi_{r}  = \frac{2}{3}\rho_{d} \bm{n} \label{eq:diffSRP}                                     \\
   & \int_{0}^{2\pi}\int_{0}^{\frac{\pi}{2}}2\rho_{s}\delta(\sin^{2}{\theta_{r}}-\sin^{2}{\theta_{i}})\delta\left(\phi_{i}-\phi_{r}\pm \pi\right) (\bm{n}^{T}\bm{v})\sin{\theta_{r}}\bm{v}\mathrm{d}\theta_{r}\mathrm{d}\phi_{r} \nonumber \\
   & = \rho_{s}\left[2(\bm{n}^{T}\bm{s})\bm{n}-\bm{s}\right] \nonumber                                                                                                                                                                 \\
   & = \rho_{s}\bm{r}_{\rm ref}\label{eq:specSRP}
\end{align}
where $\bm{r}_{\rm ref}= 2(\bm{n}^{T}\bm{s})\bm{n}-\bm{s}$ is the perfect specular direction that occurs only when $\theta_{r}= \theta_{i}$ and $\phi_{r}=\phi_{i}\pm \pi$.
A contribution from thermal re-emission of the absorbed solar flux is also included, following the standard convention used in conventional SRP modeling~\cite{mcinnes2013, wetterer2014}. Specifically, the absorbed fraction $\rho_{a}$ of the incident flux is assumed to be re-emitted instantaneously as Lambertian thermal radiation on the same side of the facet as the incident light with thermal emissivity $\kappa$. This assumption neglects the infrared emission due to heat conducted to and re-emitted from other parts of the spacecraft.
Combining this thermal emissive term with Eqs.~\eqref{eq:fSRPbrdf}--\eqref{eq:specSRP}, the conventional formulation of SRP is obtained as
\begin{align}
  \bm{f}_{\rm{SRP}} = & - \frac{{S_0}}{cr^{2}_{\rm AU}} (\bm{n}^{T}\bm{s}) A \left\{ \left( {\rho_{a}} + \rho_{d} \right)\bm{s} + \left[ \frac{2}{3}\rho_{d} + \kappa \rho_{a} + 2(\bm{n}^{T}\bm{s})\rho_{s} \right] \bm{n} \right\} \label{eq:SRP}
\end{align}
where the relation $\rho_{a}+\rho_{d}+ \rho_{s}=1$ is used.
Equation~\eqref{eq:fSRP2} cannot be analytically integrated when using the specular component of the Cook--Torrance model in Eq.~\eqref{eq:csCT} or the Ashikhmin--Shirley model in Eq.~\eqref{eq:AScs}. As a consequence, no analytical expression for the SRP can be obtained when using these high-fidelity reflection models.
For a spacecraft modeled by planar facets, the total SRP force is the sum of the forces on each facet over the illuminated and unoccluded facets.
The approximation method in this paper provides the SRP force contribution to each facet.

\subsection{Spherical Gaussians}
\begin{figure}[tb]
  \centering
  \includegraphics[width=3.25in]{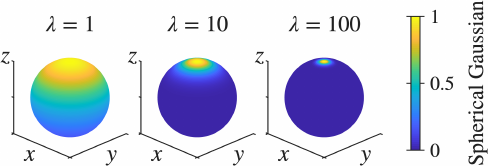}
  \caption{Examples of spherical Gaussians.}
  \label{fig:sgVis}
\end{figure}

Spherical Gaussians (SGs) have been used as a spherical basis function on a unit sphere~\cite{wang2013}.
An SG as a function of a unit vector $\bm{x}$ is defined as
\begin{align}
  \mathcal{G}\left(\bm{x}; \bm{p}, \lambda, \mu \right) = \mu e^{\lambda \left(\bm{x}^{T}\bm{p}-1\right)} \label{eq:SG}
\end{align}
where $\bm{p}$ is the lobe axis and has unit norm, $\lambda$ is the lobe sharpness, and $\mu$ is the amplitude.
The normalized SG is equivalent to the von Mises--Fisher distribution~\cite{von}.
Figure~\ref{fig:sgVis} visualizes the SGs with $\bm{p}=[0,~0,~1]^{T}$, $\mu = 1$, and various sharpnesses $\lambda = (1, 10, 100)$. The SGs reach their maximum value at the lobe direction, and the larger the sharpness, the narrower the distribution around the lobe.

Since SGs have a simple form, they have the following beneficial properties. Integrating an SG on a unit sphere, $\mathbb{S}^{2}\subset \mathbb{R}^{3}$, is analytically given as
\begin{align}
  \iint_{\mathbb{S}^{2}} \mathcal{G}(\bm{x}; \bm{p}, \lambda, \mu) \mathrm{d}\bm{x} = \mu \frac{2\pi}{\lambda}\left(1-e^{-2\lambda}\right)
\end{align}
The product of multiple SGs results in another SG as follows.
\begin{align}
  \mathcal{G}(\bm{x};\bm{p}_{1},\lambda_{1}, \mu_{1})\mathcal{G}(\bm{x};\bm{p}_{2},\lambda_{2}, \mu_{2}) = \mathcal{G}(\bm{x}; \bar{\bm{p}}_{m}, \lambda_{m}, \mu_{m}) \label{eq:sgMix}
\end{align}
where
\begin{align}
  \bm{p}_{m}       & = \lambda_{1}\bm{p}_{1}+\lambda_{2}\bm{p}_{2}                 \\
  \lambda_{m}      & = \|\bm{p}_{m}\|                                              \\
  \bar{\bm{p}}_{m} & = \frac{\bm{p}_{m}}{\|\bm{p}_{m}\|}                           \\
  \mu_{m}          & = \mu_{1}\mu_{2}e^{(\lambda_{m}-(\lambda_{1} + \lambda_{2}))} \label{eq:sgProductAmplitude}
\end{align}
These properties of SGs are used in the SRP approximation in Section~\ref{sec:approx}.

\section{Approximating SRP with the Cook--Torrance Model}\label{sec:CT}
\label{sec:approx}
An analytical approximation method is derived for the SRP formulated with the Cook--Torrance model.
Yoshimura et al.~\cite{yoshimura2022} derived an analytical approximation for the isotropic Cook--Torrance SRP by additionally replacing the SG with a piecewise linear function in half-vector space. To make the structural comparison explicit, their formulation can be written with the common physical prefactor omitted as
\begin{align}
 \mathcal{G}(\bm{h};\bm{n},\lambda_{\rm CT},1)&\approx a_{ij}+b_{ij}\bm{h}^{T}\bm{n},\qquad (\phi_h,\theta_h)\in\mathcal{D}_{ij} \\
 \bm{I}_{\rm part}&=\sum_{i=1}^{N_\phi}\sum_{j=1}^{N_\theta}\left(a_{ij}\bm{A}_{ij}+b_{ij}\bm{B}_{ij}\right) \label{eq:previousPartition}\\
 \bm{A}_{ij}&=\iint_{\mathcal{D}_{ij}}2\sin(2\theta_h)\bm{v}_h\,\mathrm{d}\theta_h\mathrm{d}\phi_h \\
 \bm{B}_{ij}&=\iint_{\mathcal{D}_{ij}}2(\bm{h}^{T}\bm{n})\sin(2\theta_h)\bm{v}_h\,\mathrm{d}\theta_h\mathrm{d}\phi_h
\end{align}
where $\bm{v}_h=2(\bm{s}^{T}\bm{h})\bm{h}-\bm{s}$, the polar angle $\theta_h$ is measured from $\bm{s}$ as in~\cite{yoshimura2022} so that the reflection Jacobian appears as $2\sin(2\theta_h)$, and $\mathcal{D}_{ij}$ are the angular integration subdomains. The coefficients $a_{ij}$ and $b_{ij}$ correspond to their local least-squares coefficients $\alpha_{ij}$ and $\beta_{ij}$. Each of the $N_\phi N_\theta$ subdomains requires two vector moments, evaluated from boundary values of trigonometric primitives.
The present method does not linearize the SG, but it keeps the exponential form, so that the properties in Eqs.~\eqref{eq:sgMix}--\eqref{eq:sgProductAmplitude} can be applied. After multiplication by the clamped-cosine SG, it requires one full-sphere vector moment per lobe:
\begin{align}
 \bm{I}_{\rm SG}=\sum_{j=1}^{N}\frac{2\pi\mu'_j}{(\lambda'_j)^2}
 \left[e^{-2\lambda'_j}(\lambda'_j+1)+\lambda'_j-1\right]\bm{p}'_j \label{eq:sgComplexityComparison}
\end{align}
Thus the summation contains $N$ vector terms ($N=1$ for Cook--Torrance and $N=5$ or $11$ for the anisotropic cases below), rather than two moments for each angular subdomain. Because each lobe contributes independently, the same construction extends directly to the Ashikhmin--Shirley model in Section~\ref{sec:AS}. The partition counts and the SG lobe count control different approximations. This structural comparison does not assert equal accuracy, and the computation time of the method in~\cite{yoshimura2022} is not measured in this paper.

\subsection{Approximation with SGs}
Since the diffuse term of the Cook--Torrance model is Lambertian and is analytically integrated as shown in Eq.~\eqref{eq:diffSRP}, this section focuses on the approximation of the specular term of the Cook--Torrance model, which is written with Eqs.~\eqref{eq:csCT} and~\eqref{eq:fSRP2} as
\begin{align}
  \bm{f}_{s, {\rm CT}}= -\frac{S_{0}}{cr^{2}_{\rm AU}}(\bm{n}^{T}\bm{s}) A\int_{0}^{2\pi}\int_{0}^{\pi/2}  \frac{D_{{\rm CT}}G_{\rm CT}(\bm{v})F_{{\rm CT}}(\bm{v})}{4(\bm{n}^{T}\bm{s})(\bm{n}^{T}\bm{v})}(\bm{n}^{T}\bm{v})\sin{\theta_{r}}{\bm{v}}\mathrm{d}\theta_{r}\mathrm{d}\phi_{r}\label{eq:fSRP}
\end{align}
where the normal vector of the facet is assumed to align with the $z$ axis. This assumption leads to the integration range $\theta_{r}\in [0, \pi/2]$ and $\phi_{r}\in [0, 2\pi]$.
This hemispherical integration range can be transformed to be spherical by replacing $\bm{n}^{T}\bm{v}$ with $\max(0,\bm{n}^{T}\bm{v})$. Note that $\max(0,\bm{n}^{T}\bm{v})$ vanishes whenever $\bm{v}$ points below the facet plane ($\bm{n}^{T}\bm{v}<0$), so the integrand identically vanishes in the lower hemisphere. Extending the integration to the full sphere is therefore purely a mathematical convenience that enables the SG-based analytical treatment, and it does not add any unphysical contribution from directions below the facet. Momentum conservation is thus preserved.
Using $M(\bm{v})=G_{\rm CT}(\bm{v})F_{\rm CT}(\bm{v})/[4(\bm{n}^{T}\bm{s})(\bm{n}^{T}\bm{v})]$ from Eq.~\eqref{eq:Mv} and the clamped cosine, Eq.~\eqref{eq:fSRP} becomes
\begin{align}
  \bm{f}_{s,{\rm CT}} & = -\frac{S_{0}}{cr^{2}_{\rm AU}}(\bm{n}^{T}\bm{s}) A\int_{0}^{2\pi}\int_{0}^{\pi} D_{\rm CT}M(\bm{v})\max{(0,\bm{n}^{T}\bm{v})}\sin{\theta_{r}}{\bm{v}}\mathrm{d}\theta_{r}\mathrm{d}\phi_{r} \label{eq:ctIntegralWithM}
\end{align}

It is known that the NDF of the Cook--Torrance model can be approximated with the following SG~\cite{wang2009}.
\begin{align}
  D_{{\rm CT}} \approx \mathcal{G}(\bm{h};\bm{n},\lambda_{\rm CT}, \mu_{\mathrm{CT}}) \label{eq:ndfSgCT}
\end{align}
where
\begin{align}
  \lambda_{\mathrm{CT}} & = \frac{2}{m^{2}}     \\
  \mu_{\mathrm{CT}}      & = \frac{1}{\pi m^{2}}
\end{align}
The remaining term of the BRDF, $M(\bm{v})$, can be approximated by its value at the perfect specular direction $\bm{r}_{\rm ref}$ defined in Eq.~\eqref{eq:specSRP}. This approximation is justified because the NDF $D_{\rm CT}$ peaks sharply around $\bm{r}_{\rm ref}$, so the integrand is dominated by values of $M$ near that direction. At this direction, $\bm{h}=\bm{n}$ and $\bm{n}^{T}\bm{r}_{\rm ref}=\bm{n}^{T}\bm{s}$, so Eq.~\eqref{eq:Mv} gives
\begin{align}
  M(\bm{r}_{\rm ref}) = \frac{G_{\rm CT}(\bm{r}_{\rm ref})F_{\rm CT}(\bm{r}_{\rm ref})}{4(\bm{n}^{T}\bm{s})^{2}}
  \qquad \bm{n}^{T}\bm{s}>0 \label{eq:MrefSubstitution}
\end{align}
The error introduced by this representative-value approximation, and a magnitude correction that compensates for it, is discussed in Section~\ref{sec:magcorrection}. Treating $M(\bm{r}_{\rm ref})$ as a constant excludes $M(\bm{v})$ from the integration as
\begin{align}
  \bm{f}_{s,{\rm CT}} & \approx -\frac{S_{0}}{cr^{2}_{\rm AU}}(\bm{n}^{T}\bm{s}) A M(\bm{r}_{\mathrm{ref}}) \int_{0}^{2\pi}\int_{0}^{\pi} \mathcal{G}(\bm{h};\bm{n},\lambda_{\mathrm{CT}}, \mu_{\mathrm{CT}})\max{(0,\bm{n}^{T}\bm{v})}\sin{\theta_{r}}{\bm{v}}\mathrm{d}\theta_{r}\mathrm{d}\phi_{r} \label{eq:fApprox}
\end{align}
The SG in Eq.~\eqref{eq:fApprox} depends on the bisector $\bm{h}$ rather than the integration variable $\bm{v}$.
For fixed $\bm{s}$, the reflection map and its solid-angle Jacobian are
\begin{align}
  \bm{v}=\bm{\psi}(\bm{h}) & =2(\bm{s}^{T}\bm{h})\bm{h}-\bm{s} \\
  J(\bm{h})=\frac{\mathrm{d}\omega_{v}}{\mathrm{d}\omega_{h}} & =4|\bm{s}^{T}\bm{h}| \label{eq:warpJacobian}
\end{align}
Following Wang et al.~\cite{wang2009}, the warped SG preserves its peak and scales its angular area by the Jacobian evaluated at the lobe center, $\bm{h}=\bm{n}$.
At a fixed relative amplitude $q\in(0,1)$, the SG cap area is $A_{q}(\lambda)=-2\pi\log q/\lambda$, provided the cap does not cover the sphere.
The condition $A_{q}(\lambda_{\rm w})=J(\bm{n})A_{q}(\lambda_{\rm CT})$ thus gives $\lambda_{\rm w}=\lambda_{\rm CT}/[4|\bm{n}^{T}\bm{s}|]$ and
\begin{align}
  \mathcal{G}(\bm{h};\bm{n},\lambda_{\rm CT},\mu_{\rm CT}) \approx \mathcal{G}\left(\bm{v};\bm{r}_{\rm ref},\frac{\lambda_{\rm CT}}{4|\bm{n}^{T}\bm{s}|},\mu_{\rm CT}\right) \label{eq:ndfApprox}
\end{align}
This isotropic approximation neglects the anisotropic distortion of the warped lobe, especially near grazing incidence.
The sharpness is constant during integration, and positive incidence $\bm{n}^{T}\bm{s}>0$ is required.

Furthermore, the clamped cosine term, $\max{(0, \bm{n}^{T}\bm{v})}$, is approximated by a single SG~\cite{wang2009} as
\begin{align}
  \max{(0,\bm{n}^{T}\bm{v})} \approx \mathcal{G}(\bm{v};\bm{n}, \lambda_{c}, \mu_{c}) \label{eq:ccApprox}
\end{align}
where $\lambda_{c} = 2.133$ and $\mu_{c}= 1.170$ are the values reported in~\cite{wang2009}, obtained by minimizing the error between the SG and the clamped cosine function.
The multiplication of two SGs yields a single SG, and consequently Eq.~\eqref{eq:fApprox} is written as
\begin{align}
  \bm{f}_{s,{\rm CT}} & \approx -\frac{S_{0}}{cr^{2}_{\rm AU}}(\bm{n}^{T}\bm{s}) A M(\bm{r}_{\mathrm{ ref}}) \iint \mathcal{G}\left(\bm{v};\bm{r}_{\mathrm{ref}},\frac{\lambda_{\mathrm{CT}}}{4|\bm{n}^{T}\bm{s}|}, \mu_{\mathrm{CT}}\right) \mathcal{G}(\bm{v};\bm{n},\lambda_{c},\mu_{c}) \sin{\theta_{r}}\bm{v}\mathrm{d}\theta_{r}\mathrm{d}\phi_{r} \\
                      & = -\frac{S_{0}}{cr^{2}_{\rm AU}}(\bm{n}^{T}\bm{s}) A M(\bm{r}_{\mathrm{ ref}})  \iint \mathcal{G}(\bm{v};\bar{\bm{p}}_{m,{\rm CT}},\lambda_{m,{\rm CT}}, \mu_{m,{\rm CT}}) \sin{\theta_{r}}\bm{v} \mathrm{d}\theta_{r}\mathrm{d}\phi_{r} \label{eq:fcsApprox}
\end{align}
where
\begin{align}
  \bm{p}_{m,{\rm CT}}       & = \frac{\lambda_{\rm CT}}{4|\bm{n}^{T}\bm{s}|}\bm{r}_{\rm ref}+\lambda_{c}\bm{n}                                                             \\
  \bar{\bm{p}}_{m,{\rm CT}} & = \frac{\bm{p}_{m,{\rm CT}}}{\|\bm{p}_{m,{\rm CT}}\|}                                                                                        \\
  \lambda_{m,{\rm CT}}      & = \|\bm{p}_{m,{\rm CT}}\|                                                                                                                    \\
  \mu_{m,{\rm CT}}          & = \mu_{c}\mu_{\rm{CT}}\exp{\left[\lambda_{m,{\rm CT}} - \left(\frac{\lambda_{\mathrm{CT}}}{4|\bm{n}^{T}\bm{s}|} + \lambda_{c}\right)\right]}
  \label{eq:SGmixCT}
\end{align}
The integration range of Eq.~\eqref{eq:fcsApprox} is spherical. To evaluate the SG vector integral alone, choose a rotated integration frame with its $z$ axis along $\bar{\bm{p}}_{m,{\rm CT}}$.
In this frame, the azimuthal integrals of the $x$ and $y$ components vanish, giving
\begin{align}
  &\int_{0}^{2\pi}\int_{0}^{\pi} \mathcal{G}(\bm{v};\bar{\bm{p}}_{m,{\rm CT}},\lambda_{m,{\rm CT}},\mu_{m,{\rm CT}}) \sin{\theta_{r}}\bm{v}\mathrm{d}\theta_{r}\mathrm{d}\phi_{r} \nonumber\\
                      &\quad = \begin{bmatrix}
                            0 \\
                            0 \\
                            \frac{2\pi\mu_{m,{\rm CT}}e^{-2\lambda_{m,{\rm CT}}}}{\lambda_{m,{\rm CT}}^{2}}\left[\lambda_{m,{\rm CT}} + 1+e^{2\lambda_{m,{\rm CT}}}(\lambda_{m,{\rm CT}}-1)\right]
                          \end{bmatrix}\label{eq:srpAnalytic}
\end{align}
Rotating this vector back to the facet-fixed frame and restoring the prefactor in Eq.~\eqref{eq:fcsApprox}, the specular part of SRP using the Cook--Torrance model is written as
\begin{align}
  \bm{f}_{s,{\rm CT}} & \approx -\frac{S_{0}}{cr^{2}_{\mathrm{AU}}} (\bm{n}^{T}\bm{s}) M(\bm{r}_{\mathrm{ref}}) A \left\{\frac{2\pi\mu_{m,{\rm CT}}}{\lambda_{m,{\rm CT}}^{2}}\left[\lambda_{m,{\rm CT}} - 1 +e^{-2\lambda_{m,{\rm CT}}}(\lambda_{m,{\rm CT}} + 1)\right]\right\} \bar{\bm{p}}_{m,{\rm CT}} \\
                      & = -\frac{S_{0}}{cr^{2}_{\rm AU}} (\bm{n}^{T}\bm{s}) M(\bm{r}_{\rm ref}) A \Lambda \bar{\bm{p}}_{m,{\rm CT}}\label{eq:approx0}
\end{align}
where
\begin{align}
  \bar{\bm{p}}_{m,{\rm CT}} & = \frac{1}{\lambda_{m,{\rm CT}}} \left( \frac{\lambda_{\rm CT}}{4|\bm{n}^{T}\bm{s}|}\bm{r}_{\rm ref}+\lambda_{c}\bm{n}\right) \label{eq:pBar}       \\
  \Lambda                   & =  \frac{2\pi\mu_{m,{\rm CT}}}{\lambda_{m,{\rm CT}}^{2}}\left[\lambda_{m,{\rm CT}} - 1 +e^{-2\lambda_{m,{\rm CT}}}(\lambda_{m,{\rm CT}} + 1)\right]
\end{align}
Equation~\eqref{eq:pBar} indicates that the SRP acts along the combined direction of $\bm{r}_{\rm ref}$ and $\bm{n}$, which analytically shows how the SRP based on the Cook--Torrance model differs from the one based on the perfect specular model. This result is physically consistent with the isotropic nature of the Cook--Torrance model, which makes the SRP invariant under rotation of the facet about the $z$ axis.

\subsection{Magnitude Correction}\label{sec:magcorrection}
The derived formulation in Eq.~\eqref{eq:approx0} includes an error due to the remaining term $M(\bm{r}_{\rm ref})$. As mentioned above, this remaining term is represented by the value at $\bm{r}_{\rm ref}$. This approximation is well accepted in the field of computer graphics, and its error is usually ignored.
However, for SRP, the error due to $M(\bm{r}_{\rm ref})$ becomes large as the roughness increases, which yields an error of more than 10\% in the SRP magnitude.
Thus, the remaining term is corrected to reduce the approximation error. Although the parameters of the correction term are determined numerically, the resulting correction substantially reduces the approximation error even when the roughness parameter is large.

The correction term $c'$ is introduced and is also described with a SG as
\begin{align}
  c' = \mathcal{G}(\bm{s};\bm{n}, \lambda', \mu') \label{eq:cprime}
\end{align}
A spherical Gaussian is chosen for the correction so that the corrected expression remains in the same family of functions as the SG approximation of the NDF, which keeps the overall formulation compact and analytically tractable.
This term is multiplied to the remaining term as
\begin{align}
  M'(\bm{r}_{\rm ref}) = c'M(\bm{r}_{\rm ref})
\end{align}
The amplitude $\mu'$ is anchored at normal incidence ($\theta_{i}=0^{\circ}$), so that the corrected approximation reproduces the exact SRP at this reference angle, where the integrand is most concentrated and the representative-value approximation is most reliable.
At normal incidence, $\bm{s}=\bm{n}$ makes the SG exponent vanish, so Eq.~\eqref{eq:cprime} reduces to $c'|_{\theta_{i}=0}=\mu'$. Therefore $\mu'$ can be obtained directly as the ratio between the exact SRP magnitude and the uncorrected approximation at $\theta_{i}=0$, both of which depend on the model parameters $m$ (through the NDF) and $F_{0}$ (through the Fresnel term); $\mu'$ is thus implicitly a function of $(m, F_{0})$.
The sharpness $\lambda'$ in Eq.~\eqref{eq:cprime} is determined by numerically minimizing the least squares error between the approximated and the true SRP magnitudes over the incident angle range $\theta_{i}\in[0, \pi/2]$, and therefore also depends implicitly on $(m, F_{0})$. Both parameters are evaluated once per material (i.e., per $(m, F_{0})$ pair) in a preprocessing step, and the optimization is inexpensive because the cost function is a scalar integral over a single variable $\theta_{i}$.
The approximation is consequently described as follows.
\begin{align}
  \bm{f}_{s, {\rm CT}} & \approx -\frac{S_{0}}{cr^{2}_{\rm AU}} (\bm{n}^{T}\bm{s}) M'(\bm{r}_{\rm ref}) A \Lambda \bar{\bm{p}}_{m,{\rm CT}} \label{eq:analyticalSRPct}
\end{align}

\begin{figure}[tb]
  \centering
  \includegraphics[width=3.25in]{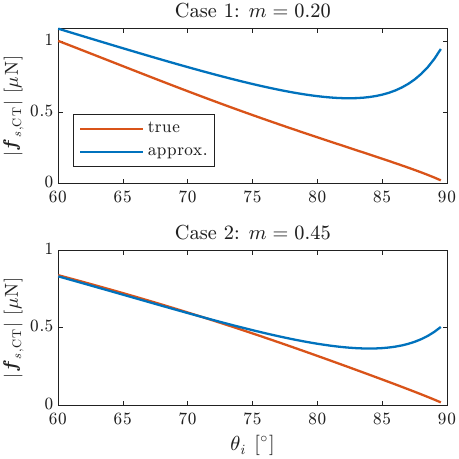}
  \caption{Cook--Torrance specular force magnitude near grazing incidence for a $1\,\mathrm{m}^2$ facet at $1\,\mathrm{au}$. }
  \label{fig:ctGrazing}
\end{figure}

The magnitude correction is evaluated for a $1~\mathrm{m} \times 1~\mathrm{m}$ flat plate for $m=0.20$ and $0.45$ at a Sun--spacecraft distance of $1~\mathrm{au}$.
The correction improves the magnitude agreement over moderate incidence angles, but it does not remove the grazing-incidence limitation illustrated in Fig.~\ref{fig:ctGrazing}.
When $\theta_{i}$ approaches $90^{\circ}$, known as the grazing angle, the approximation error becomes large.
The exact reflected force tends to zero as the illuminated projected area vanishes. The approximate expression also contains the incident cosine, but this factor is cancelled by $M(\bm{r}_{\rm ref})\propto(\bm{n}^{T}\bm{s})^{-2}$ and by the warped sharpness.
As $\bm{n}^{T}\bm{s}\to0$, $\lambda_{\rm CT}/[4|\bm{n}^{T}\bm{s}|]\to\infty$, so that $\lambda_{m,{\rm CT}}-\lambda_{\rm CT}/[4|\bm{n}^{T}\bm{s}|]\to0$, $\mu_{m,{\rm CT}}\to\mu_{c}\mu_{\rm CT}e^{-\lambda_{c}}$, and $\Lambda\to2\pi\mu_{m,{\rm CT}}/\lambda_{m,{\rm CT}}$.
With $G_{\rm CT}(\bm{r}_{\rm ref})=1$, $F_{\rm CT}(\bm{r}_{\rm ref})\to1$, and $2\pi\mu_{\rm CT}/\lambda_{\rm CT}=1$, Eq.~\eqref{eq:analyticalSRPct} has the finite grazing limit
\begin{align}
  \lim_{\bm{n}^{T}\bm{s}\to0}\|\bm{f}_{s,{\rm CT}}\|
  =\frac{S_{0}A}{cr_{\rm AU}^{2}}\,\mu_{c}e^{-\lambda_{c}}\,\mu'e^{-\lambda'}
  \approx0.139\,\frac{S_{0}A}{cr_{\rm AU}^{2}}\,\mu'e^{-\lambda'} \label{eq:ctGrazingLimit}
\end{align}
where $\mu'e^{-\lambda'}$ is the correction term at $\theta_{i}=90^{\circ}$. This limit is independent of the roughness $m$ except through the correction term, so the incident cosine alone does not guarantee the correct limiting force.
Figure~\ref{fig:ctGrazing} shows the specular-force magnitude over $60^{\circ}\leq\theta_{i}<90^{\circ}$ for the two prescribed roughnesses. The numerical force decreases toward zero, whereas the corrected approximation approaches the limit in Eq.~\eqref{eq:ctGrazingLimit}, about $1.0~\mu\mathrm{N}$ for $m=0.20$ and $0.5~\mu\mathrm{N}$ for $m=0.45$.
The approximation is applied only to illuminated facets; large errors at incident angles just below $90^{\circ}$ must be assessed over the operating range of the intended dynamics application.

\section{Approximating SRP with the Ashikhmin--Shirley Model}\label{sec:AS}
The SG-based construction developed for the isotropic Cook--Torrance model in Section~\ref{sec:CT} is extended in this section to the anisotropic Ashikhmin--Shirley model. The directional dependence of the Ashikhmin--Shirley NDF prevents its representation by a single SG, so the NDF is approximated as a superposition of a small number of SGs. Because the product and the integral of SGs both admit closed-form expressions, this superposition keeps the SRP analytically integrable, and the resulting closed-form expression has the same linear-combination-of-SGs structure as the Cook--Torrance case, with each term contributing a single vector to the SRP.

\subsection{Diffuse Reflection}
Although the diffuse term of the Ashikhmin--Shirley model is non-Lambertian, its SRP can be analytically obtained. The diffuse term of the BRDF in Eq.~\eqref{eq:AScd} is divided into two terms as
\begin{align}
  c_{d,{\rm AS}} = c_{d_{1}}c_{d_{2}}
\end{align}
where
\begin{align}
  c_{d_{1}} & = \frac{28\rho_{d}}{23 \pi} \left(1-F_{0}\right) \left[1-\left(1-\frac{\bm{n}^{T}\bm{s}}{2}\right)^{5}\right] \\
  c_{d_{2}} & = \left[1-\left(1-\frac{\bm{n}^{T}\bm{v}}{2}\right)^{5}\right]
\end{align}
Since the first term $c_{d_{1}}$ is irrelevant to integration with respect to the reference vector $\bm{v}$, it is excluded from the integration. The diffuse part of the SRP can be integrated as follows.
\begin{align}
  \bm{f}_{d,{\rm AS}} & = -\frac{S_{0}}{cr^{2}_{\rm AU}} (\bm{n}^{T}\bm{s}) A c_{d_{1}} \int_{0}^{2\pi}\int_{0}^{\pi/2} (\bm{n}^{T}\bm{v})c_{d_{2}}\sin{\theta_{r}}\bm{v}\mathrm{d}\theta_{r}\mathrm{d}\phi_{r} \label{eq:SRPd} \\
                      & =  -\frac{S_{0}}{cr^{2}_{\rm AU}} (\bm{n}^{T}\bm{s}) A c_{d_{1}} \begin{bmatrix}
                                                                                           0 \\
                                                                                           0 \\
                                                                                           \frac{1573}{2688} \pi
                                                                                         \end{bmatrix}
\end{align}
It is noted that, as shown in Fig.~\ref{fig:facet}, the integration in Eq.~\eqref{eq:SRPd} assumes the facet-fixed frame with the normal vector along the $z$ axis, resulting in the SRP along the $z$ direction.
The $x$ and $y$ components vanish because their azimuthal factors integrate to zero over $[0,2\pi]$.
For the $z$ component, the substitution $u=\cos\theta_r$ gives
\begin{align}
  &2\pi\int_0^{\pi/2}\cos^2\theta_r
  \left[1-\left(1-\frac{\cos\theta_r}{2}\right)^5\right]\sin\theta_r\,\mathrm{d}\theta_r \nonumber\\
  &\quad=2\pi\int_0^1 u^2\left[1-\left(1-\frac{u}{2}\right)^5\right]\mathrm{d}u \nonumber\\
  &\quad=2\pi\sum_{k=1}^5\frac{(-1)^{k+1}\binom{5}{k}}{2^k(k+3)}
  =\frac{1573\pi}{2688} \label{eq:asDiffuseIntegral}
\end{align}

\subsection{Specular reflection}
The specular part of SRP with the Ashikhmin--Shirley model is described using Eqs.~\eqref{eq:AScs} and~\eqref{eq:fSRP2} as
\begin{align}
  \bm{f}_{s,{\rm AS}} & = -\frac{S_{0}}{cr^{2}_{\rm AU}}(\bm{n}^{T}\bm{s}) A\int_{0}^{2\pi}\int_{0}^{\pi/2} (\bm{n}^{T}\bm{v})c_{s,{\rm AS}}\sin{\theta}_{r}\bm{v}\mathrm{d}\theta_{r}\mathrm{d}\phi_{r} \label{eq:fsAS}
\end{align}
Similar to the approximation of the Cook--Torrance model, the approximation of the specular term of the Ashikhmin--Shirley model focuses on the NDF $D_{\rm AS}(\bm{h})$.
In Section~\ref{sec:CT}, the NDF of the Cook--Torrance model is approximated with a single SG, while that of the Ashikhmin--Shirley model is approximated with multiple SGs because of anisotropy.
The specular term of the BRDF in Eq.~\eqref{eq:AScs} is rewritten as
\begin{align}
  c_{s,{\rm AS}} = M_{{\rm AS}}(\bm{v})D_{{\rm AS}}(\bm{h})
\end{align}
where $M_{{\rm AS}}(\bm{v})$ is the remaining term of the specular term except for the NDF. That is,
\begin{align}
  M_{{\rm AS}}(\bm{v}) =\frac{\sqrt{\left(n_{u}+1\right)\left(n_{v}+1\right)}}{8\pi} \frac{F_{\rm AS}}{(\bm{v}^{T}\bm{h}){\rm max}(\bm{n}^{T}\bm{s},\bm{n}^{T}\bm{v})}
\end{align}
The remaining term $M_{\rm AS}(\bm{v})$ is approximated with a representative value in the perfect specular direction $\bm{r}_{\rm ref}$, similar to the Cook--Torrance model.
Note that, when $\bm{n}^{T}\bm{s}\rightarrow 0$, the denominator of $M_{\rm AS}(\bm{r}_{\rm ref})$ is also close to zero, leading to divergence.
Thus, the remaining term is modified as follows.
\begin{align}
  M'_{{\rm AS}}(\bm{v}) = \frac{\sqrt{\left(n_{u}+1\right)\left(n_{v}+1\right)}}{8\pi} \frac{F_{{\rm AS}}}{(\bm{v}^{T}\bm{h}){\rm max}(\cos{\alpha},\bm{n}^{T}\bm{v})}
\end{align}
where $0<\alpha<90^{\circ}$ controls the denominator floor.
At the representative direction, $\bm{v}=\bm{r}_{\rm ref}$ and $\bm{h}=\bm{n}$, its effect relative to the unmodified factor is
\begin{align}
  \frac{M'_{\rm AS}(\bm{r}_{\rm ref})}{M_{\rm AS}(\bm{r}_{\rm ref})}
  =\frac{\cos\theta_i}{\max(\cos\alpha,\cos\theta_i)} \label{eq:asAlphaEffect}
\end{align}
The modification leaves this factor unchanged for $\theta_i\leq\alpha$ and reduces it above that angle.
A small $\alpha$ can suppress the force excessively, whereas a large $\alpha$ weakens the suppression near grazing incidence.
With all SG parameters fixed, changing $\alpha$ from $\alpha_0$ scales the entire approximate specular force vector by
\begin{align}
  \widehat{\bm{f}}_{s,{\rm AS}}(\alpha)=
  \frac{\max(\cos\alpha_0,\cos\theta_i)}{\max(\cos\alpha,\cos\theta_i)}
  \widehat{\bm{f}}_{s,{\rm AS}}(\alpha_0) \label{eq:asAlphaScaling}
\end{align}
For example, at $\theta_i\geq60^{\circ}$, choosing $\alpha=20^{\circ}$ or $60^{\circ}$ instead of $40^{\circ}$ changes the approximate magnitude by $-18.5\%$ or $+53.2\%$, respectively.
These are changes in the approximation and not error percentages relative to numerical integration. Their effect on the error depends on the material and incident direction.
Consequently, $\alpha$ must be checked against numerical SRP over the intended material and incident direction range. The verification uses $\alpha=40^{\circ}$ for both material cases.

\begin{figure}[tb]
  \centering
  \includegraphics[width=6cm]{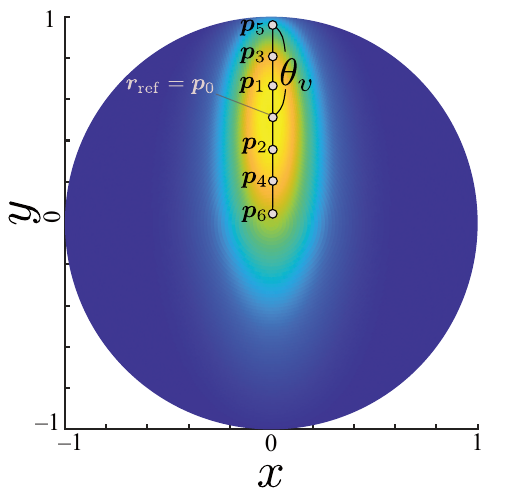}
  \caption{Lobe axis locations of SGs for Case 3.}
  \label{fig:lobeAxes}
\end{figure}

The NDF approximation by multiple SGs can be described as
\begin{align}
  D_{{\rm AS}}(\bm{h}) \approx \sum_{j=0}^{N-1}\mathcal{G}(\bm{v};\bm{p}_{j},\lambda_{j},\mu_{j}) \label{eq:NDFmixture}
\end{align}
where $N$ is the number of SGs used for the approximation.
The lobe axes $\bm{p}_{j}$ are determined by a similar method to that in~\cite{wang2009}. Their reflected directions for Case 3 are illustrated in Fig.~\ref{fig:lobeAxes}.
The axes are first located in the half-vector space along the broader material direction of the Ashikhmin--Shirley NDF and are then mapped to the reflection-direction space by spherical warping.
Let $n_{\rm broad}=\min(n_u,n_v)$ and $n_{\rm narrow}=\max(n_u,n_v)$, and let $\bm{u}_{\rm broad}$ be the material axis corresponding to $n_{\rm broad}$.
For an odd number of lobes $N=2K+1$, define the signed indices $k_j=(0,1,-1,2,-2,\ldots,K,-K)$.
The half-vector axes and their reflected directions are written as
\begin{align}
  \bar{\bm{h}}_j &= \cos(k_j\gamma)\bm{n}+\sin(k_j\gamma)\bm{u}_{\rm broad} \\
  \bm{p}_j &= 2(\bm{s}^{T}\bar{\bm{h}}_j)\bar{\bm{h}}_j-\bm{s} \label{eq:asLobeAxes} \\
  \gamma &= \frac{\theta_v}{N+1}
  \qquad \theta_v=\arccos(\epsilon_v^{1/n_{\rm broad}}) \label{eq:asLobeSpacing}
\end{align}
Thus, $\bar{\bm{h}}_0=\bm{n}$ and $\bm{p}_0=\bm{r}_{\rm ref}$.
In Eq.~\eqref{eq:asLobeSpacing}, $\gamma$ is the interval in half-vector space, and at normal incidence the reflected axes are separated by $2\gamma$.
The material parameter $\epsilon_v\in(0,1)$ sets a width scale through $\cos^{n_{\rm broad}}\theta_v=\epsilon_v$. The outermost half-vector axes lie at $\pm K\gamma$, inside this width scale.
The common lobe sharpness is selected by matching two normalized angular profiles at a prescribed angle $\theta_u$ from the lobe axis in reflected-direction space.
For normal incidence, a reflected-direction offset $\theta_u$ corresponds to a half-vector offset $\theta_u/2$, so the narrow-direction cross section of Eq.~\eqref{eq:Das} at this offset is $\cos^{n_{\rm narrow}}(\theta_u/2)$.
Equating it to the unit-peak SG at the same offset gives
\begin{align}
  \exp\left[\lambda_j(\cos\theta_u-1)\right] &= \cos^{n_{\rm narrow}}(\theta_u/2) \label{eq:asSharpnessMatch} \\
  \lambda_j &= \frac{n_{\rm narrow}\log[\cos(\theta_u/2)]}{\cos\theta_u-1}
  \qquad j=0,1,\ldots,N-1 \label{eq:asSharpness}
\end{align}
where $\theta_u=\arccos\epsilon_u$ and $0<\epsilon_u<1$. The parameter $\epsilon_u$ is the cosine of the matching angle and the matched amplitude is $[(1+\epsilon_u)/2]^{n_{\rm narrow}/2}$.
This condition equates the two normalized profiles at one angle and does not impose equality of the entire SG mixture and the NDF. The same sharpness is used for all lobes and incident directions as an approximation.
The lobe amplitudes are
\begin{align}
  \mu_j=\begin{cases}
    D_{\rm AS}(\bar{\bm{h}}_j)/C_\mu & \bm{s}^{T}\bar{\bm{h}}_j>0 \\
    0 & \bm{s}^{T}\bar{\bm{h}}_j\leq0
  \end{cases} \label{eq:asLobeAmplitude}
\end{align}
where $C_\mu>0$ is a common amplitude scale for each material case. The positive-dot-product condition ensures that $\bar{\bm{h}}_j$ is the bisector of $\bm{s}$ and $\bm{p}_j$; otherwise that lobe is inactive.
Other lobes also contribute at $\bm{p}_j$, so assigning an amplitude to an individual lobe does not force the mixture to equal the NDF at that axis.
The parameters $N$, $\epsilon_u$, $\epsilon_v$, and $C_\mu$ are prescribed for each verification case in Section~\ref{sec:verify}. The NDF plots in Section~\ref{sec:verify} are normalized to unit peak for visualization, whereas the SRP calculation uses the amplitudes of Eq.~\eqref{eq:asLobeAmplitude}.

In the same way as for the approximation of the Cook--Torrance model, the dot product $\bm{n}^{T}\bm{v}$ in Eq.~\eqref{eq:fsAS} is approximated with the clamped cosine term, ${\rm max}(0,\bm{n}^{T}\bm{v})$, allowing the conversion of the hemispherical integration range to a spherical one.
The clamped cosine term is approximated with a SG as shown in Eq.~\eqref{eq:ccApprox}.
The SRP formulation using the SG mixture and the clamped cosine approximation is written as
\begin{align}
  \bm{f}_{s,{\rm AS}} & = -\frac{S_{0}}{cr^{2}_{\rm AU}}(\bm{n}^{T}\bm{s}) A\int_{0}^{2\pi}\int_{0}^{\pi} {\rm max}(0,\bm{n}^{T}\bm{v})c_{s,{\rm AS}}\sin{\theta}_{r}\bm{v}\mathrm{d}\theta_{r}\mathrm{d}\phi_{r}                                                                                                                                  \\
                      & \approx -\frac{S_{0}}{cr^{2}_{\rm AU}} (\bm{n}^{T}\bm{s}) AM'_{\rm AS}(\bm{r}_{\rm ref})\int_{0}^{2\pi}\int_{0}^{\pi} \mathcal{G}(\bm{v};\bm{n},\lambda_{c},\mu_{c})\sum_{j=0}^{N-1}\mathcal{G}(\bm{v};\bm{p}_{j},\lambda_{j},\mu_{j}) \sin{\theta_{r}}\bm{v}\mathrm{d}\theta_{r}\mathrm{d}\phi_{r} \label{eq:sgMixApprox} \\
                      & = -\frac{S_{0}}{cr^{2}_{\rm AU}} (\bm{n}^{T}\bm{s}) AM'_{\rm AS}(\bm{r}_{\rm ref}) \sum_{j=0}^{N-1} \bm{f}_{j}
\end{align}
where
\begin{align}
  \bm{f}_{j} = \int_{0}^{2\pi}\int_{0}^{\pi} \mathcal{G}(\bm{v};\bm{n},\lambda_{c},\mu_{c})\mathcal{G}(\bm{v};\bm{p}_{j},\lambda_{j},\mu_{j})\sin{\theta_{r}}\bm{v}\mathrm{d}\theta_{r}\mathrm{d}\phi_{r}\label{eq:fj}
\end{align}

The remainder of this section focuses on $\bm{f}_{j}$, an individual term in the summation.
The two SGs in the integrand in Eq.~\eqref{eq:fj} are rewritten as a single SG using the SG mixture in Eq.~\eqref{eq:sgMix} as
\begin{align}
  \bm{f}_{j} = \int_{0}^{2\pi}\int_{0}^{\pi} \mathcal{G}(\bm{v};\bm{p}'_{j},\lambda'_{j},\mu'_{j}) \sin{\theta_{r}}\bm{v}\mathrm{d}\theta_{r}\mathrm{d}\phi_{r} \label{eq:srpMixed}
\end{align}
Equation~\eqref{eq:srpMixed} can be integrated following the same procedure used for the Cook--Torrance model, yielding
\begin{align}
  \bm{f}_{j} = \frac{2\pi\mu'_{j}e^{-2\lambda'_{j}}}{\lambda_{j}'^{2}}\left[\lambda'_{j} + 1+e^{2\lambda'_{j}}(\lambda'_{j}-1)\right] \bm{p}'_{j}
\end{align}
Consequently, the specular SRP is approximated as follows.
\begin{align}
  \bm{f}_{{\rm SRP},s} \approx -\frac{S_{0}}{cr^{2}_{\rm AU}} (\bm{n}^{T}\bm{s}) A M'_{\rm AS}(\bm{r}_{\rm ref})\sum_{j=0}^{N-1} \frac{2\pi\mu'_{j}e^{-2\lambda'_{j}}}{\lambda_{j}'^{2}}\left[\lambda'_{j} + 1+e^{2\lambda'_{j}}(\lambda'_{j}-1)\right] \bm{p}'_{j} \label{eq:specularSRPapprox}
\end{align}
This is a generalized approximation that extends the SRP approximation of the Cook--Torrance model in Eq.~\eqref{eq:analyticalSRPct} by employing multiple spherical Gaussians.

\section{Numerical Verification}\label{sec:verify}
All force calculations in this section use the same $1~\mathrm{m}\times1~\mathrm{m}$ flat plate at a Sun--spacecraft distance of $1~\mathrm{au}$.
The plate lies in the facet-fixed $x$--$y$ plane with $\bm{n}=[0,0,1]^T$, and the reported force components are expressed in this frame.
Since the diffuse part of the SRP is analytical for both models, the specular part of the SRP is verified by comparing the approximation results with the true values obtained by numerical integration.
The approximation is verified sequentially for three targets: 1) the normal distribution function, 2) the clamped-cosine term, and 3) the specular SRP.
Table~\ref{tab:condition} lists two cases of optical properties for each reflection model. The strong anisotropy case $(n_{u},n_{v})=(800, 50)$ for the Ashikhmin--Shirley model corresponds to a brushed aluminum material experimentally fitted in~\cite{kurt2010}. The Fresnel reflectance is set to $F_{0}= 0.5$ for all cases.
\begin{table}[tb]
  \begin{center}
    \caption{Test cases}\label{tab:condition}
    \begin{tabular}{lcc}
      \hline \hline
      Cook--Torrance model     & roughness $m$                \\ \hline
      Case 1                   & 0.20                       & \\
      Case 2                   & 0.45                       & \\
      \hline
      Ashikhmin--Shirley model & anisotropy $(n_{u},n_{v})$   \\ \hline
      Case 3                   & (100, 50)                  & \\
      Case 4                   & (800, 50)                  & \\
      \hline \hline
    \end{tabular}
  \end{center}
\end{table}

\subsection{SRP with the Cook--Torrance model}
\begin{figure}[tb]
  \centering
  \includegraphics[width=3.25in]{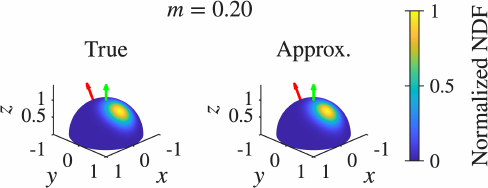}
  \caption{Visualization of normalized NDF approximation of the Cook--Torrance model with $m=0.20$. Red and green arrows denote the Sun vector and the facet normal, respectively.}
  \label{fig:ndfCT1}
\end{figure}

\begin{figure}[tb]
  \centering
  \includegraphics[width=3.25in]{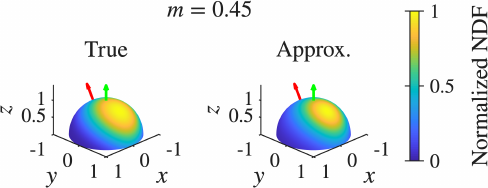}
  \caption{Visualization of normalized NDF approximation of the Cook--Torrance model with $m=0.45$. Arrow colors have the same meaning as in Fig.~\ref{fig:ndfCT1}.}
  \label{fig:ndfCT2}
\end{figure}

\begin{figure}[tb]
  \centering
  \includegraphics[width=3.25in]{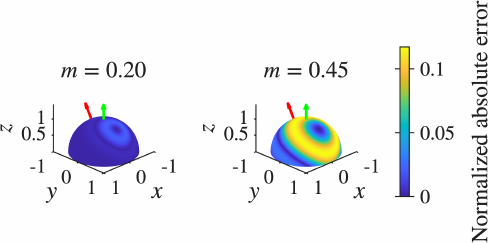}
  \caption{Normalized absolute errors of NDF approximation of the Cook--Torrance model. Arrow colors have the same meaning as in Fig.~\ref{fig:ndfCT1}.}
  \label{fig:ndfCTabsErr}
\end{figure}

In this example, the Sun vector is set to $(\phi_{i},\theta_{i}) = (-90, 30)^{\circ}$ and the normal vector is along the $z$ axis. Figures~\ref{fig:ndfCT1} and~\ref{fig:ndfCT2} respectively visualize the approximation of NDFs for two cases of roughness, $m=0.20$ and $0.45$. For each case, both the exact and approximate NDFs are normalized by the maximum value of the exact NDF. The approximate NDF in these figures is the half-vector-space SG of Eq.~\eqref{eq:ndfSgCT}, evaluated at the bisector of each plotted direction, and the spherical warping of Eq.~\eqref{eq:ndfApprox} is applied only in the SRP derivation.
The distribution is well approximated for $m=0.20$, whereas the approximated distribution for $m=0.45$ is slightly narrower than the true one.
Figure~\ref{fig:ndfCTabsErr} shows the normalized absolute errors of the NDF approximation.
The maximum errors occur almost along the direction of $\bm{r}_{\rm ref}$.
This approximation is justified because the SRP exhibits its highest values in regions where the NDF reaches maximum values.

The SG mixture with the clamped cosine approximation in Eqs.~\eqref{eq:ccApprox} and~\eqref{eq:fcsApprox} is verified.
To exclude the error due to the remaining term $M(\bm{r}_{\rm ref})$, the SG mixture combining the clamped cosine approximation and an arbitrary SG $\mathcal{G}(\bm{v},\bm{p}_{1},\lambda_{1},\mu_{1})$ is examined, which is written as
\begin{align}
  \iint {\rm max}(0, \bm{n}^{T}\bm{v})\mathcal{G}(\bm{v},\bm{p}_{1},\lambda_{1},\mu_{1})\sin{\theta_{r}}\bm{v}\mathrm{d}\theta_{r}\mathrm{d}\phi_{r} & \approx  \iint \mathcal{G}(\bm{v};\bm{n},\lambda_{c},\mu_{c})\mathcal{G}(\bm{v};\bm{p}_{1},\lambda_{1},\mu_{1})\sin{\theta_{r}}\bm{v}\mathrm{d}\theta_{r}\mathrm{d}\phi_{r} \label{eq:ccApprox111} \\
                                                                                                                                                     & =  \iint \mathcal{G}(\bm{v};\bm{p}'_{1},\lambda'_{1},\mu'_{1}) \sin{\theta_{r}}\bm{v}\mathrm{d}\theta_{r}\mathrm{d}\phi_{r}\label{eq:ccApprox2}                                                    \\
                                                                                                                                                     & = \frac{2\pi\mu'_{1}e^{-2\lambda'_{1}}}{\lambda_{1}'^{2}}\left[\lambda'_{1} + 1+e^{2\lambda'_{1}}(\lambda'_{1}-1)\right] \bm{p}'_{1}
\end{align}
where Eq.~\eqref{eq:ccApprox111} is a general expression of the SG mixture with the clamped cosine approximation. The left-hand side of Eq.~\eqref{eq:ccApprox111} is numerically calculated as a true value.
Changing the angle of the lobe axis, $\theta_{p}=\arccos{(\bm{n}^{T}\bm{p_{1}})}$, in the $x-z$ plane, the approximation errors of $\lambda_{1}=\frac{2}{m^{2}}=50.00$ for Case~1 and $\lambda_{1} = 9.88$ for Case~2 are examined.

\begin{figure}[!htbp]
  \centering
  \includegraphics[width=3.25in]{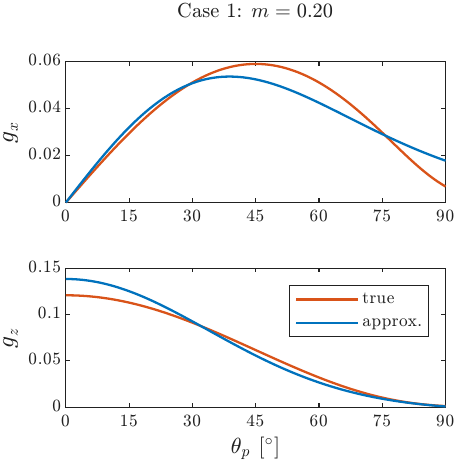}
  \caption{Spherical Gaussian mixture with clamped cosine for Case 1. }
  \label{fig:sgMixCCct1}
\end{figure}

\begin{figure}[!htbp]
  \centering
  \includegraphics[width=3.25in]{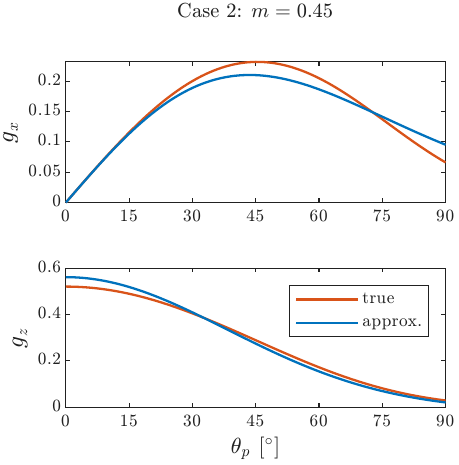}
  \caption{Spherical Gaussian mixture with clamped cosine for Case 2.}
  \label{fig:sgMixCCct2}
\end{figure}

Figures~\ref{fig:sgMixCCct1} and~\ref{fig:sgMixCCct2} show the approximation results of the SG mixture with clamped cosine for Cases 1 and 2, respectively. Since the lobe axis is varied in the $x-z$ plane, the $y$ component is identically zero and is omitted. The two panels show the in-plane $x$ component and the normal $z$ component. The clamped cosine approximation implemented with a single SG keeps the relative error below roughly a few percent for incidence angles away from the grazing region. Its acceptability depends on the accuracy required by the intended application.
Although multiple SGs in the clamped cosine approximation would yield higher accuracy, it increases the analytical complexity. Thus, this paper adopts a single SG approximation for the clamped cosine.

\begin{figure}[!htbp]
  \centering
  \includegraphics[width=3.25in]{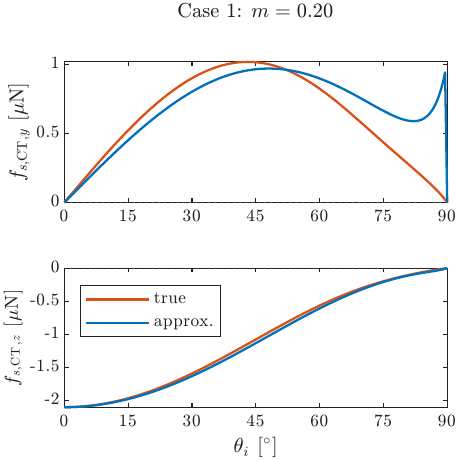}
  \caption{Specular SRP approximation for Case 1.}
  \label{fig:ctSRP1}
\end{figure}

By varying the Sun direction $\theta_{i}$, the specular SRP including the magnitude correction in Eq.~\eqref{eq:analyticalSRPct} is verified. Since the Cook--Torrance model is an isotropic model, the azimuth angle of the Sun direction is set to $\phi_{i}=90^{\circ}$ without loss of generality.
Figure~\ref{fig:ctSRP1} compares the true and approximate values for the specular SRP with the Cook--Torrance model. Because the incident Sun direction is varied in the $y-z$ plane, the $x$ component is identically zero by symmetry and is omitted. The two panels show the nonzero components: the in-plane component along the projection of the Sun vector onto the facet, $f_{s,{\rm CT},y}$, and the normal component, $f_{s,{\rm CT},z}$.
The approximation along the $z$ axis is accurate, while the one along the $y$ axis deviates as $\theta_{i}$ becomes large. This deviation arises because the bisector $\bm{h}$ moves further from the facet normal $\bm{n}$ at large incidence angles, which degrades the SG approximation of the NDF in directions away from the specular peak and introduces a tangential error in the SRP.
The magnitude of the specular SRP is also comparable when $\theta_{i}$ is less than $60^{\circ}$.

\begin{figure}[!htbp]
  \centering
  \includegraphics[width=3.25in]{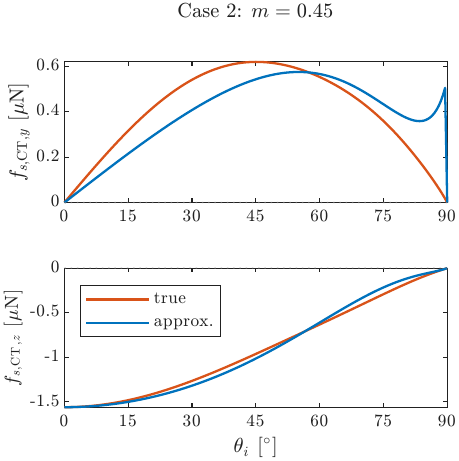}
  \caption{Specular SRP approximation for Case 2. The panels show the nonzero in-plane $y$ and normal $z$ components.}
  \label{fig:ctSRP2}
\end{figure}

Figure~\ref{fig:ctSRP2} shows the approximation result for Case 2 using the same two nonzero components. Similarly to the previous case, the SRP along the $y$ axis has a relatively larger error than the normal component. The normal component agrees more closely than the tangential component, and the force magnitude deviation increases near grazing incidence.

\FloatBarrier

\subsection{SRP with the Ashikhmin--Shirley model}
\begin{figure}[tb]
  \includegraphics[width=3.25in]{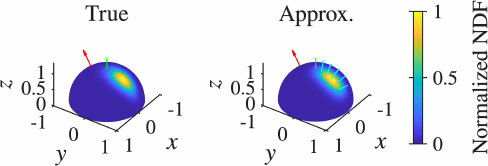}
  \centering
  \caption{Normal distribution function approximation for Case 3: true (left) and approximation (right). Red, green, and cyan arrows denote the Sun vector, the facet normal, and the SG lobe axes, respectively.}
  \label{fig:ndf01}
\end{figure}

\begin{figure}[tb]
  \centering
  \includegraphics[width=0.5\textwidth]{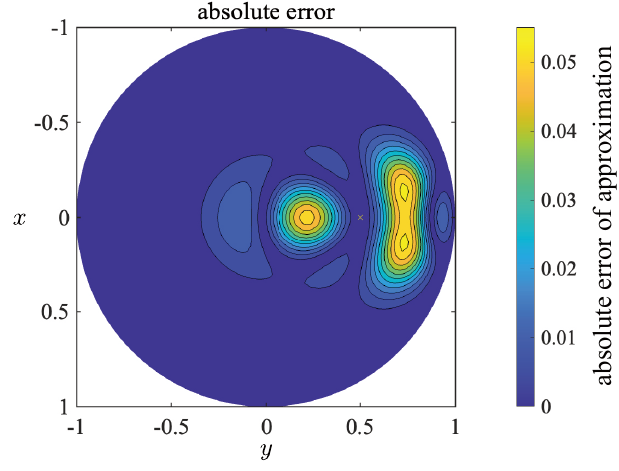}
  \caption{Absolute error of the NDF approximation for Case 3.}
  \label{fig:ndf01AbsErr}
\end{figure}

\begin{figure}[tb]
  \centering
  \includegraphics[width=3.25in]{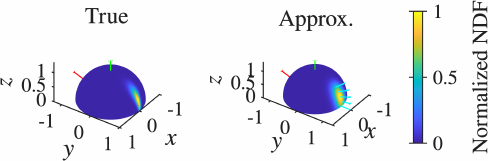}
  \caption{Normal distribution function approximation for Case 3 at grazing angle: true (left) and approximation (right). Arrow colors have the same meaning as in Fig.~\ref{fig:ndf01}.}
  \label{fig:ndf01Graze}
\end{figure}

\begin{figure}[tb]
  \centering
  \includegraphics[width=0.5\textwidth]{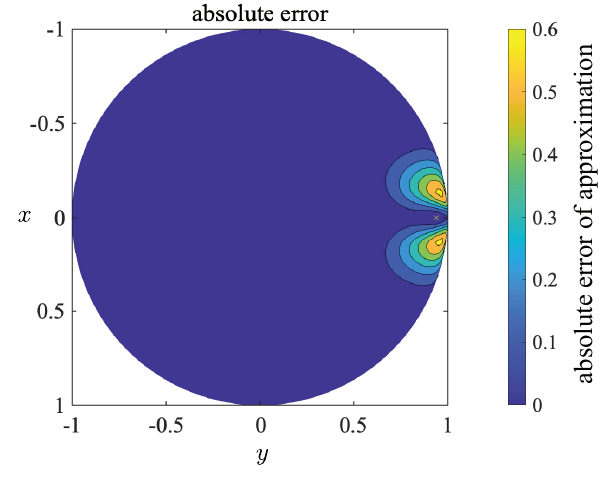}
  \caption{Absolute error of the NDF approximation for Case 3 at grazing angle.}
  \label{fig:ndf01GrazeErr}
\end{figure}

The SG parameters used for the Ashikhmin--Shirley verification are listed in Table~\ref{tab:asSgParameters}.
\begin{table}[tb]
  \centering
  \caption{SG parameters for the Ashikhmin--Shirley cases.}
  \label{tab:asSgParameters}
  \begin{tabular}{ccccc}
    \hline\hline
    Case & $N$ & $\epsilon_u$ & $\epsilon_v$ & $C_\mu$ \\
    \hline
    3 & 5 & 0.05 & 0.001 & 2.183888 \\
    4 & 11 & 0.2 & 0.0001 & 1.5635 \\
    \hline\hline
  \end{tabular}
\end{table}
The approximation of the NDF with multiple SGs in Eq.~\eqref{eq:NDFmixture} is examined. The Sun vector is set to $(\phi_{i},\theta_{i}) = (-90, 30)^{\circ}$ and the parameters $\epsilon_{u} = 0.05$ and $\epsilon_{v}=0.001$ are used for Case~3.
Figure~\ref{fig:ndf01} visualizes the true NDF and the approximation by five SGs, demonstrating that the NDF in the hemisphere is accurately captured with the SGs.
Figure~\ref{fig:ndf01AbsErr} represents the absolute error, where the perfect specular direction $\bm{r}_{\rm ref}$ is also plotted for better understanding.
Figure~\ref{fig:ndf01Graze} visualizes the NDF approximation for the Sun vector at $(\phi_{i},\theta_{i}) = (-90, 70)^{\circ}$.
In this case, the Sun vector and normal vector are close to perpendicular, i.e., the grazing angle.
Even if the optical properties $n_{u}$ and $n_{v}$ have weak anisotropy, reflection near the grazing angle represents strong anisotropy.
Figure~\ref{fig:ndf01GrazeErr} shows the absolute error at this grazing angle.

\begin{figure}[tb]
  \centering
  \includegraphics[width=3.25in]{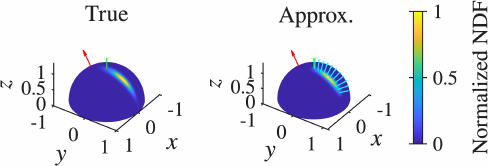}
  \caption{Normal distribution function approximation for Case 4: true (left) and approximation (right). Arrow colors have the same meaning as in Fig.~\ref{fig:ndf01}.}
  \label{fig:ndf02}
\end{figure}

\begin{figure}[tb]
  \centering
  \includegraphics[width=0.5\textwidth]{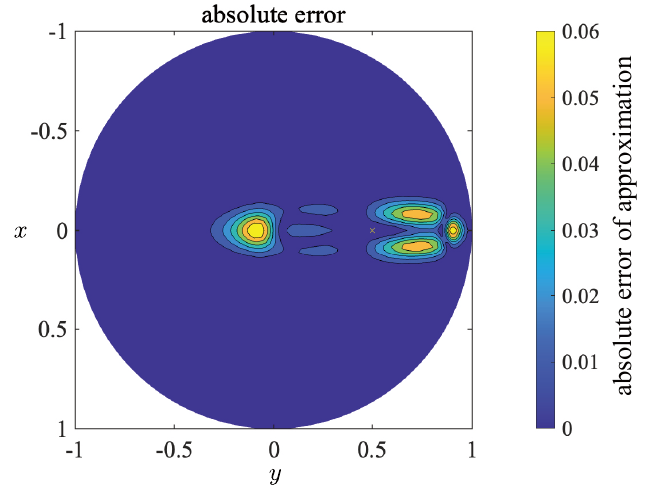}
  \caption{Absolute error of the NDF approximation for Case 4.}
  \label{fig:ndf02AbsErr}
\end{figure}

\begin{figure}[tb]
  \centering
  \includegraphics[width=3.25in]{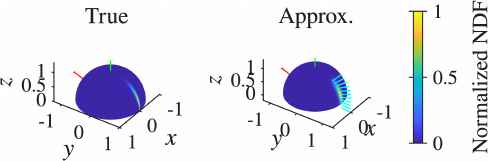}
  \caption{Normal distribution function approximation for the strong anisotropy case at grazing angle: true (left) and approximation (right).Arrow colors have the same meaning as in Fig.~\ref{fig:ndf01}.}
  \label{fig:ndf02Graze}
\end{figure}

\begin{figure}[tb]
  \centering
  \includegraphics[width=3.25in]{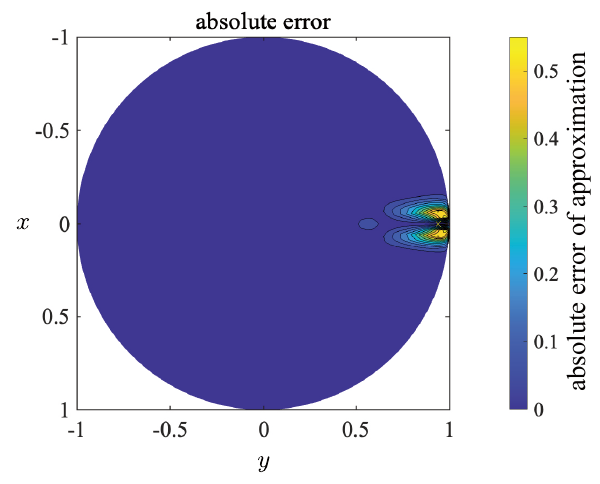}
  \caption{Absolute error of the NDF approximation for the strong anisotropy case at grazing angle.}
  \label{fig:ndf02GrazeErr}
\end{figure}

For Case 4, the parameters $\epsilon_{u} = 0.2$ and $\epsilon_{v}=0.0001$ are used and 11 SGs are used to represent the stronger anisotropy.
The visual comparison of the NDF approximation is shown in Fig.~\ref{fig:ndf02} and shows good agreement with the true value.
Figure~\ref{fig:ndf02AbsErr} shows the absolute error.
Figures~\ref{fig:ndf02Graze} and~\ref{fig:ndf02GrazeErr} show the NDF approximation and the absolute error for the Sun vector at $(\phi_{i},\theta_{i}) = (-90, 70)^{\circ}$, respectively.

\begin{figure}[tb]
  \centering
  \includegraphics[width=3.25in]{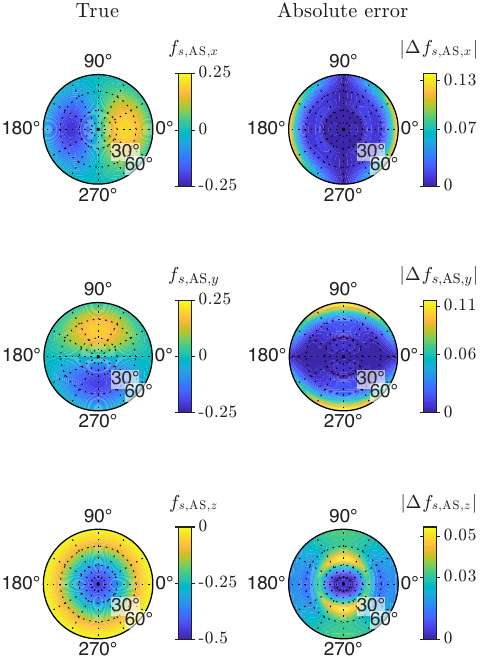}
  \caption{Specular SRP approximation for Case 3: true values (left) and absolute approximation errors (right), normalized by $S_0A/(cr_{\rm AU}^2)$. The radial coordinate is $\theta_i$ and the circumferential coordinate is $\phi_i$, both in degrees.}
  \label{fig:srpApprox01All}
\end{figure}

\begin{figure}[tb]
  \centering
  \includegraphics[width=3.25in]{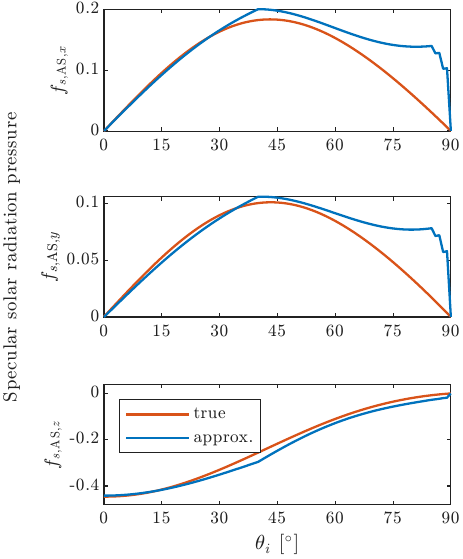}
  \caption{Sliced view of specular SRP approximation for Case 3 ($\phi_{i} = 30^{\circ}$), normalized by $S_0A/(cr_{\rm AU}^2)$.}
  \label{fig:srpApprox01}
\end{figure}

By varying the Sun direction, the specular SRP including the remaining term $M(\bm{r}_{\rm ref})$ in Eq.~\eqref{eq:specularSRPapprox} is verified.
Since the Ashikhmin--Shirley model is an anisotropic reflection model, the azimuth angle of the Sun direction is also varied over the range $\phi_{i}\in [0,360)^{\circ}$.
Both cases use $\alpha=40^{\circ}$ in the remaining term. As shown in Eq.~\eqref{eq:asAlphaEffect}, the denominator modification takes effect above $\theta_i=40^{\circ}$, and its sensitivity is described by Eq.~\eqref{eq:asAlphaScaling}.
For Case 3, Fig.~\ref{fig:srpApprox01All} shows the true specular SRP and the approximation error, where the left column represents the true values for each component and the right column represents the absolute difference between the approximate and true values. In Figs.~\ref{fig:srpApprox01All} and~\ref{fig:srpApprox01}, the forces are normalized by $S_0A/(cr_{\rm AU}^2)$, and the true values are obtained by Monte Carlo integration with $10^5$ importance samples of the half vector; they agree with the deterministic quadrature used for Table~\ref{tab:asErrorSummary} within $3\times10^{-4}$ in normalized units.
The radial coordinate represents the elevation angle of the Sun vector, $\theta_i$, and the circumferential coordinate represents the azimuth angle, $\phi_i$; both angles are labeled in degrees in the figure.
The approximation captures the dominant angular dependence of the anisotropic SRP, while localized error patterns remain, especially when $\theta_{i}$ is close to the grazing angle.
Figure~\ref{fig:srpApprox01} compares the true and approximate values at $\phi_{i}=30^{\circ}$ and is the sliced view of Fig.~\ref{fig:srpApprox01All}.
This sliced view shows the same trend and the approximation reproduces the main variation with $\theta_{i}$, whereas the error increases at large Sun angles.
The NDF approximation and the representative remaining factor both contribute to these grazing-incidence errors.
The shrinking projected area reduces the absolute force near grazing incidence, but does not by itself establish that the approximation error is negligible for a particular dynamics or control application.

The polar contour plot is shown for Case 3, and the quantitative errors for both anisotropy cases are summarized in Table~\ref{tab:asErrorSummary}.
The table uses the incident-direction grid $\theta_i=0^{\circ},3^{\circ},\ldots,87^{\circ}$ and $\phi_i=0^{\circ},5^{\circ},\ldots,360^{\circ}$, with the deterministic reference quadrature described below.
Let $K=30\times73$ denote the number of grid entries, and let $\bm{f}_k$ and $\widehat{\bm{f}}_k$ be the numerically integrated and approximate specular forces at entry $k$, respectively.
The two percentage errors with $\Delta\bm{f}_k=\widehat{\bm{f}}_k-\bm{f}_k$ are defined as
\begin{align}
  E_{z,\max} &= 100\,\frac{\max_{1\leq k\leq K}|\Delta f_{k,z}|}
  {\max_{1\leq k\leq K}|f_{k,z}|} \label{eq:asMaxNormalError}\\
  E_{\rm RMS} &= 100\,\frac{\sqrt{K^{-1}\sum_{k=1}^{K}\|\Delta\bm{f}_k\|^2}}
  {\sqrt{K^{-1}\sum_{k=1}^{K}\|\bm{f}_k\|^2}} \label{eq:asRmsVectorError}
\end{align}
These are discrete grid statistics with equal weight for each entry, including the coincident azimuthal endpoints $0^{\circ}$ and $360^{\circ}$, rather than solid angle weighted integrals.
The normal component error is normalized by the peak reference force component, and the vector error by the reference RMS magnitude.
The normal component, which is the dominant SRP component for the flat facet, has maximum errors of 12.2\% and 11.3\% of its peak magnitude for Cases 3 and 4, respectively.
The RMS vector errors are 16.0\% and 13.6\%, while the largest localized deviations occur near large Sun angles.

\begin{table}[tb]
  \centering
  \caption{Quantitative error summary for the Ashikhmin--Shirley specular SRP approximation. The normal-component maximum is normalized by the peak value of $\left|f_{s,{\rm AS},z}\right|$. The RMS vector error is normalized by the RMS value of $\|\bm{f}_{s,{\rm AS}}\|$.}
  \label{tab:asErrorSummary}
  \begin{tabular}{ccccc}
    \hline\hline
    Case & $(n_u,n_v)$ & $N$ &
    \begin{tabular}{c}Max. normal\\component [\%]\end{tabular} &
    \begin{tabular}{c}RMS\\vector [\%]\end{tabular} \\
    \hline
    3 & $(100,50)$ & 5  & 12.2 & 16.0 \\
    4 & $(800,50)$ & 11 & 11.3 & 13.6 \\
    \hline\hline
  \end{tabular}
\end{table}

\subsection{Reference integration}\label{sec:comparison}
The reference specular forces are evaluated by deterministic quadrature in half-vector space, which uses 256 Gauss--Legendre nodes in $\cos\theta_h\in[0,1]$ and 512 uniformly spaced midpoint nodes in $\phi_h\in[0,2\pi)$. The reflection map $\bm{v}=2(\bm{s}^{T}\bm{h})\bm{h}-\bm{s}$ has Jacobian $\mathrm{d}\omega_v=4(\bm{s}^{T}\bm{h})\mathrm{d}\omega_h$ on the valid domain $\bm{s}^{T}\bm{h}>0$, $\bm{n}^{T}\bm{v}>0$. For the Ashikhmin--Shirley model, the normalized half-vector sampling density is written as
\begin{align}
 p_h(\bm{h})=\frac{\sqrt{(n_u+1)(n_v+1)}}{2\pi}D_{\rm AS}(\bm{h})\qquad(\bm{n}^{T}\bm{h}>0) \label{eq:asSamplingDensity}
\end{align}
This density is defined with respect to solid angle and contains no additional factor $\bm{n}^{T}\bm{h}$. Consequently, its importance-sampling weight for reflected momentum contains $F_{\rm AS}(\bm{h})(\bm{n}^{T}\bm{v})/\max(\bm{n}^{T}\bm{s},\bm{n}^{T}\bm{v})$ on the valid domain.
Table~\ref{tab:asErrorSummary} uses this deterministic reference throughout. Refining a $128\times256$ rule to the stated rule changes the force vector, normalized by $S_0A/(cr_{\rm AU}^2)$, by at most $2.9\times10^{-5}$ over all four cases and the incident grid. Independent viewing-direction quadrature with $768\times1536$ nodes agrees within $4\times10^{-6}$ in the same units at six selected directions per case, including $\theta_i=87^{\circ}$. For both anisotropic cases, an independent $10^6$-sample Monte Carlo check at four incident angles agrees within two estimated standard errors in each component. These checks assess reference integration accuracy, separately from the much larger model-approximation errors.

\subsection{Computational cost}\label{sec:cost}
Table~\ref{tab:timing} compares calculation times on an Apple M4 Max using MATLAB R2026a. Each call evaluates the total force for a facet at $(\theta_i,\phi_i)=(45^{\circ},30^{\circ})$. Table~\ref{tab:timing} reports the median of seven warmed batches with 1000 calls per batch for SG evaluation and 10 for each numerical integrator. Quadrature nodes, NDF weights, Monte Carlo half-vector samples, and material parameters are prepared before timing; parameter fitting, random-number generation, file I/O, and plotting are excluded. The Monte Carlo timing evaluates the sample mean without a standard error calculation. All methods include the incident and analytical diffuse contributions.
\begin{table}[tb]
\centering
\caption{Calculation time comparison for a single facet and incident direction. Quadrature uses $256\times512$ nodes and Monte Carlo uses $10^5$ importance samples of the half vector.}
\label{tab:timing}
\begin{tabular}{cccc}
\hline\hline
Case & SG [$\mu$s] & Quadrature [$\mu$s] & Monte Carlo [$\mu$s]\\
\hline
1 & 1.1 & 1362 & 1127 \\
2 & 0.8 & 1219 & 917 \\
3 & 3.6 & 1272 & 902 \\
4 & 3.1 & 1135 & 626 \\
\hline\hline
\end{tabular}
\end{table}
The SG evaluation is faster than these numerical integration methods by two to three orders of magnitude. The timings measure the cost of the stated implementations and integration resolutions, and they are not comparisons at equal error tolerance.

\FloatBarrier
\section{Conclusions}\label{sec:conc}
This paper has proposed an analytical approximation of the solar radiation pressure (SRP) for high-fidelity reflection models, including the anisotropic Ashikhmin--Shirley model.
The normal distribution function (NDF) is approximated by a small number of spherical Gaussians (SGs), and the closure properties of SGs yield a force expression in which each lobe contributes a single vector term. The isotropic Cook--Torrance model is the single lobe special case.
For weak and strong anisotropy, the approximation reproduces the dominant angular dependence of the specular force with RMS vector errors of 16\% and 14\%, respectively, and it is two to three orders of magnitude faster than numerical integration.
The proposed method is intended for repeated force evaluation in orbit and attitude dynamics, where the closed form can be summed over facets and embedded in stability analysis and controller design.
Its accuracy is bounded by the SG representation of the NDF, the clamped cosine approximation, and the representative bidirectional reflectance distribution function factor with a per-material magnitude correction.
Near grazing incidence, the approximation does not vanish with the projected area and the anisotropic lobes are distorted. Thus, the errors just below $\theta_i=90^{\circ}$ are large.
For the Ashikhmin--Shirley model, the result also depends on the threshold $\alpha$ and the prescribed lobe parameters. These settings should be validated against numerical SRP over the material and incident-angle range of the intended application.

\section*{Acknowledgments}
The authors used Claude (Anthropic) to assist with language editing of the manuscript. The authors reviewed all AI-assisted content and take full responsibility for the manuscript.

\clearpage
\bibliography{asrp}

\end{document}